\documentstyle[preprint,aps]{revtex}
\input{psfig.sty}
\begin{document}
\draft
\preprint{ HU-SEFT-R-1995-15}
\title{Resurrection of the Sigma Meson}
\author{Nils A. T\"ornqvist}
\address{Research Institute for High Energy Physics, SEFT
POB 9, FIN--00014, University of Helsinki, Finland}
\author{Matts Roos}
\address{High Energy Physics Laboratory,
POB 9, FIN--00014, University of Helsinki, Finland}
\date{September 1995}
\maketitle
\begin{abstract}
It is shown from a
very general model and an analysis of data on the lightest
$0^{++}$ meson nonet
that the $f_0(980)$ and $f_0(1200)$ resonance poles are two
manifestations of the same $ s\bar s $ state.
Similarily the $a_0(980)$ and the $a_0(1450)$ are likely
to be two manifestations of the same $ q\bar q$ state.
On the other hand, the $ u\bar u+d\bar d $ state,
when unitarized and strongly distorted by hadronic mass shifts,
becomes an extremely broad Breit-Wigner-like background,
$m_{BW}=860$ MeV, $\Gamma_{BW}=880$ MeV, with its pole at $s$=
(0.158-i0.235) GeV$^2$. This we identify with the $ \sigma\ $ meson
required by models for spontaneous breaking of chiral symmetry.
\\
\medskip \noindent
\end{abstract}

\medskip 

The understanding of the $\pi\pi$ S-wave has been  controversial
for a long time. Before 1974 (see \cite{PDGold}) one believed in the
existence of a broad and light isoscalar resonance (then called $\sigma$ or
$\epsilon$ or $\eta_{0+}$).
After the two heavier resonances $f_0(980)$ and $f_0(1200-1300)$
 were established
one generally discarded the $\sigma$ assuming it could be replaced by the
two
heavier ones, which could complete a light $ q\bar q$ nonet.

The lightest scalar-isoscalar meson coupling strongly to
$\pi\pi$ is of importance in most models for spontaneous breaking
of chiral symmetry, like the linear $\sigma$ model or
the Nambu--Jona-Lasinio
model\cite{Lee}, which require a scalar meson of twice the
constituent quark
mass or $\approx$700 MeV, and a very large
$\pi\pi$ width  of $\approx$850 MeV.
This meson is crucial for our understanding
of all hadron masses. Thus
most of the nucleon mass is believed to be generated
by its coupling to
the $\sigma$, which acts like an effective Higgs-like
 boson for the hadron spectrum.  However,  the lightest
well established
mesons with the quantum numbers of the $\sigma$,
the $f_0(980)$ and $f_0(1200)$\, do not have the right properties.
They are both too narrow, $f_0(980)$ couples mainly
to $K\bar K$, and $f_0(1200)$ is too heavy.

Recently one of us\cite{NAT} showed that one can understand the data
on the
lightest scalars in a model which includes most well established
theoretical constraints: Adler zeroes as required by chiral symmetry,
all light two-pseudoscalar (PP) thresholds with flavor
symmetric couplings,
physically acceptable analyticity, and unitarity.  A unique
feature of this model is that it simultaneously describes the  whole
scalar
nonet and one obtains a good representation of a large
set of relevant data. Only six parameters,
which all have a  clear physical interpretation,  were needed:
an overall coupling constant
($\gamma=1.14$), the bare mass of the $u\bar u$ or $d\bar d$ state
($m_0=1.42$ GeV), the extra mass for a strange quark
($m_s=100$ MeV),
a cutoff parameter ($k_0=0.56$ GeV/c), an Adler zero
parameter for $K\pi$
($s_{A_{K\pi}}=-0.42$ GeV), and a phenomenological
parameter\footnote{\baselineskip 10pt One could discard the $\beta$
parameter if one also included the next group of important
thresholds or pseudoscalar ($0^{-+}$) -axial ($1^{+-}$)
 thresholds, since then the $K\bar K_{1B}+c.c.$
thresholds give a very similar contribution to the mass
matrix as $\eta\eta'$.}
enhancing the $\eta\eta '$ couplings ($\beta=1.6$).

As expected, the coupling to PP turns out\cite{NAT}
to be very large causing
the inverse propagators to have  very large imaginary
parts. Normally, this
is expected to result in large widths, but it was shown\cite{NAT}
 that because of the many nonlinear effects the $a_0(980)$
and the $f_0(980)$ naturally come out narrow.
Furthermore, the large flavor symmetry breaking in the
positions of the PP
thresholds induce large flavor symmetry breaking in the mass shifts.
This makes the physical spectrum quite distorted compared to the
simple bare spectrum, which obeys the equal spacing rule and the
OZI rule with flavor symmetric couplings.

The analysis\cite{NAT} yielded resonance parameters and
pole positions
for the four states $a_0(980)$ , $f_0(980)$ , $f_0(1200)$
and $K_0^*(1430)$ ,
close to their  conventional values\cite{PDG94}.
In addition, however, one also expects "image" poles
(sometimes called "shadow" or "companion" poles), which normally lie
far away and do not play a significant r\^ole. Recently
Morgan and Pennington\cite{Morgan} showed that for each
$ q\bar q$ state one
 expects at least one such image pole, which in principle can be used
to distinguish a $ q\bar q$ state from a meson-meson bound state.

In this letter we report a more detailed
 search for all relevant poles in the amplitudes of the  model
of \cite{NAT}. This reveals interesting and surprising
new features, which
simultaneously resolve two longstanding puzzles in meson
spectroscopy:

\begin{itemize}
\item What is the nature of the $f_0(980)$ and $f_0(1200)$?
\item Where is the long sought for $\sigma$ meson?
\end{itemize}

These  new, important features can appear when the coupling to S-wave
thresholds becomes very strong. In particular,
 two true resonance poles can emerge near the
physical region, although only one $ q\bar q$ state is present.
If the coupling is reduced one of these poles disappears
as a distant image
pole far from the physical region.

In order to explain this phenomenon in the simplest
possible terms we use two
theoretical demonstrations. The first is based on the
actual model amplitude
in \cite{NAT}, which fits the $K\pi$   S-wave data
and the $K_0^*(1430)$.
We chose the  channel with strangeness,  because there one has only
one  $s\bar d$ quark model state, whereas in the flavorless case
one has the more complicated situation of at least
two nearby resonances, the  $ u\bar u+d\bar d
$ and $ s\bar s $ , which mix in an energy dependent and
complex way.
    By increasing
the overall coupling, $\gamma$, we show how  a second pole appears.
In the second demonstration we chose a simple model for the
threshold behaviour, which still has the desired analytic properties,
and from which the poles can be found analytically. This can also
be used to demonstrate the phenomenon for the $K\bar K$ channel.

In Fig.\ 1a we show the running mass, $m_0^2+Re\Pi (s)$
and the width-like
function $-Im \Pi (s)$ for the $K\pi \to K\pi$ S-wave as was found in
\cite{NAT}. The $K\pi$
partial wave amplitude is obtained from these functions through

\begin{equation}
A(s)=-Im\Pi_{K\pi} (s)/[m_0^2+Re\Pi (s)-s +iIm \Pi (s)]\
. \label{PWA}
\end{equation}

 This fits the $K\pi$ data
well and one finds the resonance parameters  listed in the first row
of Table I.
As one increases $\gamma$  there appears,
in addition to the resonance, first a
virtual bound state and then a true bound state just below the $K\pi$
threshold (See Figs. 1b,c and Table I). Both poles, the
original $K_0^*(1430)$ and the new bound state are then
manifestations of the same $s\bar u$ state, whose bare
mass is kept at 1520 MeV.

For our  second way to demonstrate this phenomenon we
chose the form factor such
that $\Pi (s)$ takes a simple analytic form,
\begin{equation}
\Pi (s)= \bar\gamma^2s_{th}
[((s_{th}-s)s_{th})^{1/2}-s_{th}]/s\ . \label{pii}
\end{equation}

 See Fig. 2. This  still has the desired analytic form
and satisfies the dispersion relation (See discussion in Sec 2.7 of
\cite{NAT}).
The condition for the poles, $m_0^2+\Pi (s) -s =0$, now
gives an equation of
third degree. If $m_0^2=s_{th}(1+\bar \gamma^2)$ one has
one bound state
at threshold $s=s_{th}$ and (for $\bar\gamma >1/2$)
a complex conjugate pair at $s=s_{th}[\bar\gamma^2
+1/2\pm i(\bar\gamma^2-1/4)^{1/2}]$. For
$\bar\gamma^2\raise0.3ex\hbox{$>
$\kern-0.75em\raise-1.1ex\hbox{$\sim$}} 1$
one has a running mass (cf. Fig.~2) quite similar to the
one in Fig~1c for the
$K\pi$ threshold and to the actual one fitting the data at
the $K\bar K$
threshold (Fig.~2b and 9a of
 ref. \cite{NAT}). If $\bar\gamma>1$ the phase shift passes through
90$^\circ$ at
$s=\bar\gamma^2s_{th}=m_{BW}^2$. This simplified model
is also similar to the
actual situation in \cite{NAT} for the $ s\bar s $ channel
and the $K\bar K$
threshold. The $\bar\gamma$ of the simplified model is comparable
in magnitude to the $\gamma$ used in \cite{NAT}, such that
with $\gamma=\bar
\gamma$ both models give similar $Im\Pi_{K\bar K} (s)$ near the
$K\bar K$ threshold  for $ s\bar s $. Thus the fact
that $\gamma=1.14 >1 $
actually shows that the real world is not too far from our simplified
model.   Using $\bar \gamma =1.14$ one would predict
 $(Re s_{pole})^{1/2}=
2m_K(\bar\gamma^2+1/2)^{1/2}= 1329$ MeV and
$m_{BW}= 2m_K \bar\gamma =1129$ MeV,
which is not far from what was actually obtained
for the $f_0(1200)$ : 1202, and 1186
respectively.
In reality, of course, other thresholds and mixing
with $ u\bar u+d\bar d $
complicate the picture.  For example
the $f_0(980)$ and the $a_0(980)$ are probably
unstable virtual states (i.e., lying on the
second sheet, and  not on the
third sheet although two thresholds are open at the pole positions).

The situation is also similar for the
$a_0(980)$ and $a_0(1450)$ in the I=1 channel,
although now the Clebsch-Gordan coefficient reduces
the effective $\bar\gamma$
by $1/\sqrt 2$. However, the fact that the  $\pi\eta$
channel is already
open at the $K\bar K$ threshold helps in creating
a similar situation of a
running mass rising fast enough after threshold. Therefore,
one can expect
a repetition of the phenomenon, such that there could exist a second
manifestation of the I=1 state,  somewhere in the 1.5
GeV region, in addition to $a_0(980)$.
This could be the $a_0(1450)$ seen by the
Crystal Barrel\cite{amsler}.
And as we shall see below, the model of \cite{NAT} actually has
image poles near this mass, one of which in an improved model
and fit could
emerge as the $a_0(1450)$.
On the other hand, in the strange channel, there is only
one important channel open, the $K\pi$ with a
Clebsch-Gordan coefficient
reducing the coupling compared to $s\bar s- K\bar K$
by $(3/4)^{1/2} $.
This, together with the fact that the $K\pi$
threshold involves two unequal
mass mesons, implies that the resonance doubling
phenomenon does not appear
in the strange sector.

We now look for the actual pole positions in the model of
\cite{NAT}, and list the significant ones in Table II.
Four of these, which are near the physical region,
were already given in
\cite{NAT}: the $f_0(980)$ , $f_0(1200)$ ,
$a_0(980)$ and $K_0^*(1430)$. However, we now find  three
new poles. (There are of course more image poles,
which we do not list,
since these are very far from the
physical region.) Note that all the poles in Table II
are manifestations of {\it the same nonet}.
In  \cite{NAT} the Breit-Wigner parameters of
the $\sigma$ meson were also
given, but it was not specified which pole
should be associated with it.

We now unambiguously find that the $f_0(980)$ and
the $f_0(1200)$ are two
manifestations of the same $ s\bar s $ state.
The dominant pole in $ u\bar u+d\bar d $ is the
first pole in Table II at
$s=(0.158-i0.235) $GeV$^2$, which gives rise to
a very broad (880 MeV)
Breit-Wigner-like background with $m_{BW}=860$ MeV.
One can convince oneself that this is the right
conclusion by decoupling the
two channels $ s\bar s $ and $ u\bar u+d\bar d $.
 This can be done within the model, maintaining
unitarity etc., by sending $m_0$ or $m_0+2m_s$
(gradually) to infinity.
The $\sigma$ pole remains always almost at the same position as in
Table II even when $m_s \to \infty$,
while there is no trace of $f_0(980)$ nor $f_0(1200)$\
in $ u\bar u+d\bar d $. The two latter poles remain, however,
in the $ s\bar s $ channel
even when this is completely decoupled from $ u\bar u+d\bar d $.
 The $f_0(980)$ remains near
the $K\bar K\ $ threshold whereas $f_0(1200)$
is shifted to somewhat higher values.

A posteriori, this result is natural also in the light
of the mixing angles,
$\delta_S$, found for the physical states (see Table II).
At the $\sigma$ pole, as well as for energies
$\;\raise0.3ex\hbox{$<$\kern-0.75em\raise-1.1ex\hbox{$\sim$}}\; 900$
 MeV $\delta_S$
is small along the real $s$ axis.
Thus the $\pi\pi$ amplitudes below this energy are
dominated by the $\sigma$
 and only slightly perturbed by the $ s\bar s $ and $K\bar K$
channels. The $f_0(980)$ and the near
octet\footnote{\baselineskip 10pt The near octet nature
of this state is also supported by the small
branching ratio of 0.02 to
$\eta\eta$ fond by GAMS2000\cite{alde},
since the $8-\eta\eta$ coupling nearly
vanishes for the conventional pseudoscalar mixing angle.
}
 $f_0(1200)$  owe their existence to the
$ s\bar s $$\to K\bar K$ channel dynamics and have a
comparatively small mixing with
the $ \sigma $, also evident from the rather small mixing angle
$\delta_S$ of these states.

In conclusion, the $\sigma$ meson exists in the $ u\bar u+d\bar d $
 channel and we tentatively
name it $f_0(600)$. This is the scalar meson
required by models for dynamical
breaking of chiral symmetry. It has the right mass
and width and large $\pi\pi$
coupling, thus dominating $\pi\pi$ scattering below 900 MeV.
The very large width of the $\sigma$
explains why it has been
difficult to find in the data, without a
sufficiently sophisticated model.

 The $f_0(980)$ and $f_0(1200)$ are both
manifestations of the same
$ s\bar s $ quark model state.
  Similarily we believe that the $a_0(980)$ and the $a_0(1450)$
are two manifestations of the I=1 $ q\bar q$ quark model
state. In the $s\bar u$ and $ u\bar u+d\bar d $
systems the image
poles of the $K_0^*(1430)$ and $\sigma$, respectively,
are sufficiently far from the physical region
and therefore do not give rise to
additional resonances.  We emphasize again
that all the states discussed in this paper
are manifestations of the same quark model
nonet, which naturally can be assumed to be the $^3P_0$ nonet.
When unitarized, the $^3P_0$ naive quark model spectrum
is strongly distorted, and results   not in
4, but  in 6 different physical resonances of different isospin.

One could argue that the two states $f_0(980)$
and $a_0(980)$ are a kind of
$K\bar K$ molecules, since these have a large component of $K\bar K$
in their wave function. However, the dynamics of
these states is quite
different from that of normal two-hadron bound states. In
particular, it is very different from the
hyperfine interaction suggested by
Weinstein and Isgur \cite{wein}. If one
wants to consider them as molecules,
it is the $K\bar K \to s\bar s \to K\bar K$
interaction which  creates their
binding energy. Although they may spend most of
their time as $K\bar K$ they
owe their existence to the $ s\bar s $ state. And in general,
it is not obvious which of the two states,
$f_0(980)$ or $f_0(1200)$ , would be the extra state
and which is $ q\bar q$ ,
since one can well imagine situations (with different
$m_0$ and $\gamma$) where either one of these is removed
to being a distant image pole. Therefore, one should
rather consider both as
two different manifestations of the same $ q\bar q$ state.

\null

\begin{figure}
\null
\vfill
\caption{(a) The running mass and
-Im$\Pi (s)$ which fits the $K\pi$ S wave
data. In (b) and (c) the overall coupling,
$\gamma$, is increased
whereby first a virtual bound state (b)
and then a bound state (c) appears
below the $K\pi$ theshold in addition
to the $K^*_0$ resonance whose
$90^\circ$ mass gets shifted to slightly lower energies.
See text and Table I.}
\end{figure}
\eject
\setcounter{figure}{0}
\begin{figure}
\null
\vfill
\caption{(b) As in fig 1a the running mass and -Im$\Pi (s)$
but with the overall coupling, $\gamma$, is increased to 1.34
whereby first a virtual bound state  appears
below the $K\pi$ theshold in addition
to the $K^*_0$ resonance whose
$90^\circ$ mass gets shifted to slightly
lower energies. See text and Table I.}
\end{figure}
\eject

\setcounter{figure}{0}
\begin{figure}
\null
\vfill
\caption{(c) The running mass and -Im$\Pi (s)$
but with the overall coupling, $\gamma$, is increased to 1.53
whereby  a bound state  appears
below the $K\pi$ theshold in addition to the $K^*_0$ resonance whose
$90^\circ$ mass gets shifted to slightly
lower energies. See text and Table I.}
\end{figure}
\eject
\null

\begin{figure}
\null
\vfill
\caption{The running mass and $-Im\Pi (s)$
for the simple model function (2) and the $K\bar K$ threshold
using $\bar \gamma=1.14$. Choosing $m_0$ such that there is
 a bound state at the theshold, there is another
resonance  as seen from the second crossing of $s$ with the
running mass. This crossing point gives
the $90^\circ$ or Breit-Wigner mass
of the second state. }
\end{figure}

\eject
{\squeezetable
\begin{table}
\caption{ \baselineskip 10pt
The 90$^\circ $ "Breit-Wigner" mass and width together with the
pole positions of the $K_0^*(1430)$ resonance in units of MeV.
The first row
shows the result from the actual fit to data, while the
second and third
rows  show how the resonance parameters are shifted
as one increases
$\gamma $. At the same time, first a virtual bound state
and then a true bound
state appears when $\gamma $ is increased.
 The bare $ s\bar s $ mass is 1520 MeV
throughout. See also Fig. 1.}
\begin{tabular}{|c|c|c|c|c|l|}
\hline
$\gamma$ & $m_{BW}$ &$\Gamma_{BW}$&
$m_{pole}=Re(s_{pole})^{1/2}$ & $-Im
(s_{pole})/m_{pole}$ & Comment\\
\hline
1.14     & 1349    & 498       &   1441 & 270 &
No nearby image pole\\
1.34     & 1260 & 780&1460 & 404 &
A virtual bound state appears at 548 MeV \\
1.53     & 1100 & 1100  & 1496 & 441 &
 A bound state appears at 632 MeV\\
\hline
\end{tabular}
\end{table}}

{\squeezetable
\begin{table}
\caption{ \baselineskip 9pt
Poles in the S-wave  $ PP \to PP $ amplitudes [3]
The first resonance is the $ \sigma\ $
which we name $f_0(600)$. The two following are both
manifestations of the same $s\bar s$ state.
The $f_0(980)$ and $a_0(980)$ have no
approximate Breit Wigner-like description,
and the $\Gamma_{BW}$ given for
$a_0(980)$ is rather the peak width.
The two last entries are image poles of the
$a_0(980)$ , one of which in an improved fit
could represent the $a_0(1450)$.
The $f_0(1200)$ and $K_0^*(1430)$ poles appear simultaneously
on two sheets since the $\eta\eta$ and the
$K\eta$ couplings, respectively,
 nearly vanish. The mixing angle $\delta_S $ for the
$f_0(600)$ or $\sigma$  is with respect to
$u\bar u +d\bar d$, while for the two heavier
$f_0$'s it is with respect to $s\bar s$. }
\begin{tabular}{|c|c|c|c|c|c|c|c|l|}
\hline
resonance&$m_{BW}$&$\Gamma_{BW}$&$\delta_{S,BW}$&$m_{pole}
$&$-Im (s_{pole})/m_{pole}$&$\delta_{S,pole}$&Sheet&Comment\\
\hline
$f_0(600)$&860&880&$(-9+i8.5)^\circ$&397&590&$
(-3.4+i1.5)^\circ $&II&The $\sigma$ meson. Near
$ u\bar u+d\bar d $ state\\
$f_0(980)$&-&-&-&1006&33.7&$(0.4+i39)^\circ$&II&
First near $ s\bar s $ state\\
$f_0(1200)$&1186&360&$(-32+i1)^\circ$&1202&338&$
(-36+i2)^\circ$&III,V&Second
near $ s\bar s $ state\\
$K_0^*(1430)$&1349&498&-&1441&320&-&II,III&The $s\bar d$ state\\
$a_0(980)$&987&$\approx$100&-&1084&270&-&II&First I=1 state\\
$a_0(1450)?$&-&-&-&1566&578&-&III&Image pole to $a_0(980)$ \\
$a_0(1450)?$&-&-&-&1748&690&-&V&Image pole to $a_0(980)$ \\
\hline
\end{tabular}
\end{table}
}

\end{document}